\documentclass{JHEP3}
\usepackage{graphicx,psfrag,epsfig,fancyvrb,makeidx,latexsym,amssymb,amsmath}
\DefineVerbatimEnvironment{code}{Verbatim}{fontsize=\small}
\def\lsim{\mathrel{\vcenter{\hbox{$<$}\nointerlineskip\hbox{$\sim$}}}}
\def\gsim{\mathrel{\vcenter{\hbox{$>$}\nointerlineskip\hbox{$\sim$}}}}
\def\mET{E_T \hspace{-1.2em}/\;\:}

\makeindex
\preprint{RECAPP-HRI-2012-009}
\title	{Phenomenology of Light Sneutrino Dark Matter in cMSSM/mSUGRA with Inverse Seesaw}
\author{P. S. Bhupal Dev,$^{a,b}$ Subhadeep Mondal,$^{c}$ Biswarup Mukhopadhyaya$^{d}$ and Sourov Roy$^{c}$\\
$^{a}$ Maryland Center for Fundamental Physics, Department of Physics, University of Maryland, College Park, MD 20742, USA.\\
$^b$ Consortium for Fundamental Physics, School of Physics and Astronomy, University of Manchester, Manchester, M13 9PL, United Kingdom.\\
$^c$ Department of Theoretical Physics,
Indian Association for the Cultivation of Science,
2A \& 2B Raja S.C. Mullick Road, Kolkata 700032, India.\\
$^d$ Regional Centre for Accelerator-based Particle Physics, Harish-Chandra Research Institute, Chhatnag
Road, Jhusi, Allahabad 211019, India. 
\\ 
\email{Bhupal.Dev@hep.manchester.ac.uk, tpsm2@iacs.res.in, biswarup@hri.res.in, tpsr@iacs.res.in}
}
\abstract{We study the possibility of a light Dark Matter (DM) within a constrained Minimal Supersymmetric Standard Model (cMSSM) framework augmented by a SM singlet-pair sector to account for the non-zero neutrino masses by inverse seesaw mechanism. Working within a `hybrid' scenario with the MSSM sector fixed at high scale and the singlet neutrino sector at low scale, we find that, contrary to the case of the usual cMSSM where the neutralino DM cannot be very light, we can have a light sneutrino DM with mass below 100 GeV satisfying all the current experimental constraints from cosmology, collider as well as low-energy experiments. We also note that the supersymmetric inverse seesaw mechanism with sneutrino as the lightest supersymmetric partner can have enhanced same-sign dilepton final states with large $\mET$~ coming from the gluino- and squark-pair as well as the squark-gluino associated productions and their cascade decay through charginos.
We present a collider study for the same-sign dilepton+jets+$\mET$~ signal in this scenario and propose some distinctions with the usual cMSSM. We also comment on the implications of such a light DM scenario on the invisible decay width of an 125 GeV Higgs boson.}
\begin{document}
\maketitle
\tableofcontents
\section{Introduction}
While the existence of Dark Matter (DM) in our universe is by now well-established from various astrophysical and cosmological observations~\cite{dmobserve}, its nature and properties are still unknown. 
Many experimental efforts are under way~\cite{dmobserve,pdgdm} to identify the DM candidate in various direct detection experiments through its scattering 
off different nuclei as well as from indirect detection through its annihilation products. Since no particle in the Standard Model (SM) can qualify as the DM candidate which is required to be dominantly `cold'~\cite{dmobserve}, these studies will also be sensitive probes of  physics beyond the SM and supplement the new physics search at the Large Hadron Collider (LHC). 

Many extensions of the SM indeed require the introduction of new particles, some of which could be DM candidates. Perhaps the most popular candidate for particle DM is the lightest supersymmetric particle (LSP) in $R$-parity conserving supersymmetric (SUSY) models~\cite{jungman}. In the Minimal Supersymmetric extension of the SM (MSSM), the lightest neutralino is the usual  
DM candidate, as the other viable candidate, namely, the scalar superpartner 
of the left-handed (LH) neutrino, is strongly disfavored by a combination of relic density, 
direct detection and invisible decay width of the SM $Z$-boson constraints~\cite{hebb}. More specifically, 
the unsuppressed coupling of the LH sneutrino to the SM $Z$-boson leads to a large cross-section for elastic scattering with the target nuclei in direct detection devices~\cite{212}; such cross-sections are already ruled out experimentally over almost the entire viable mass range~\cite{xenon}. By the same argument, their large 
$Z$-coupling leads to far too rapid annihilation for the LH sneutrinos and too small a relic density compared to the Wilkinson Microwave Anisotropy Probe (WMAP)-measured value~\cite{wmap}. One could make them very light (of order GeV)~\cite{211} in order to suppress the annihilation rate 
as well as to evade the direct detection bounds due to the limited sensitivity of the experiments at low masses. However, a very light sneutrino is 
excluded by the measurement of the $Z$-boson invisible decay width at the Large Electron-Positron (LEP) collider~\cite{lep}.

On the other hand, the recent data from three direct detection experiments, namely, DArk MAtter (DAMA)~\cite{dama}, Coherent Germanium Neutrino Technology (CoGeNT)~\cite{cogent} and Cryogenic Rare Event Search with Superconducting Thermometers 
(CRESST)~\cite{cresst}, have suggested the hints of a light DM with mass in the $10-100$ GeV range and cross section in the range $10^{-3}-10^{-6}$ pb for elastic scattering off nucleons. 
Although there is no unanimity among these results and several other direct detection experiments~\cite{xenon,xenonold,cdms,cdmsedel,edel,zeplin,simple} do not see any such positive hints of a particle DM, it has certainly generated considerable curiosity in a light DM scenario which can explain some/all of the hints of positive detection while being consistent with the null results from other direct detection experiments~\cite{dmtheo}. Therefore, it may not be premature to examine some beyond SM scenarios accommodating a light DM candidate in case any of these positive hints are confirmed in near future.  

If we assume gaugino mass unification in the MSSM, the LEP collider searches of SUSY put a lower bound on the 
lightest neutralino mass of around 50 GeV~\cite{pdg1}, and the recent LHC data push this bound to more than about 200 GeV~\cite{pdg2,beskidt}. Even if we do not assume gaugino mass unification, one could derive a lower limit on the neutralino LSP mass of $\sim 20$ GeV~\cite{hooper} just  
requiring the observed DM relic density, together with the LEP constraints on chargino and slepton masses\footnote{For a recent discussion on how to avoid this lower bound in a non-universal MSSM parameter space, see e.g., Ref.~\cite{light} and references therein.}. Therefore, if the DM indeed turns out to be very light as suggested by some of the recent experiments~\cite{dama,cogent,cresst}, we need to go beyond the universal scenario. 
Since the MSSM anyway cannot be a complete theory and needs to be extended to accommodate the observed small neutrino masses~\cite{pdgvu}, it would be interesting to see if these extensions can also provide a viable light DM candidate while satisfying both the collider and relic density constraints as well as other low-energy constraints in the leptonic sector. 

A simple way to understand the smallness of neutrino masses is by the seesaw mechanism~\cite{seesaw}. The canonical type-I seesaw requires the addition of 
one set of heavy SM singlet Majorana fermions to the particle content of the SM, and the smallness of the light neutrino masses are due to the heaviness of the Majorana neutrinos. In its supersymmetric version, the lightest superpartner of the singlet neutrinos with a small admixture of the left-sneutrino
can be the LSP with some fine adjustment of parameters or in an extension of 
the MSSM gauge group (see e.g., Ref.~\cite{mazumdar} and references therein). In such a case, it can be a viable light DM candidate.  

In this paper, we focus on the supersymmetric version of a different seesaw scenario, namely, the inverse seesaw~\cite{inverse} where one adds two sets of SM singlet fermions, one Dirac $N$ and one Majorana $S$ per family, to generate the small neutrino masses.
These models have three lepton-number carrying electrically-neutral fermions per family, namely, $(\nu_L, N^c, S)$. If a linear combination of the super-partners 
of these fields turns out to be the LSP, then it could be a scalar DM candidate. Current literature on the subject discusses several variations of such models, e.g., (i) within the framework of MSSM~\cite{valle}, (ii) in Next-to-Minimal Supersymmetric SM (NMSSM)~\cite{kang2011}, (iii) with extended gauge symmetry $SU(2)_L\times U(1)_Y\times U(1)_{B-L}$~\cite{khalil} so that the seesaw mass matrix arises from a $B-L$ gauge symmetry~\cite{khalil2}\footnote{Another class of models~\cite{michaux} uses global $B-L$ symmetry to restrict the inverse seesaw matrix to the desired form. Yet another recent work~\cite{leptino} uses a gauged $U(1)$-extension with only renormalizable operators, but adding pairs of fermions with fractional lepton number.}, (iv) with Supersymmetric Left-Right gauge group $SU(2)_L\times SU(2)_R\times U(1)_{B-L}$~\cite{adcm} where the inverse seesaw texture arises naturally even at TeV scale while preserving the gauge coupling unification~\cite{model}. 

In the present work, we take a hybrid approach similar to that in Ref.~\cite{valle}, i.e., a low-energy input for the $SU(2)_L$-singlet neutrino sector and for the lepton-number violating soft SUSY-breaking sector while a top-down approach for the MSSM particle spectrum, without necessarily imposing any features of a specific Grand Unified Theory (GUT)-based model. 
Our goal is to examine if such a minimal Supergravity (mSUGRA) scenario with inverse seesaw can give a light DM candidate satisfying all the existing 
cosmological, collider as well as low-energy constraints. If so, we ask ourselves what the collider signals for such a scenario are and how to distinguish it from a typical 
constrained MSSM (cMSSM) scenario for a similar squark-gluino spectrum at the LHC. In particular, since the sneutrino mass is not directly related to the gaugino masses, and there are additional unknown parameters in the sneutrino mass matrix, we expect the lightest sneutrino to be allowed to have masses in the few GeV range without being in conflict with the collider bounds on gluino and chargino  masses. Then it remains to be checked whether this lightest sneutrino eigenstate has the right admixture of left- and singlet-sneutrino flavors to reproduce the observed relic density while satisfying the constraints from direct and indirect detection experiments as well as from other low-energy sectors. 
In fact, we find that in contrast with the usual cMSSM scenario~\cite{beskidt}, we 
can have the sneutrino LSP mass in the few GeV range while being consistent with the SUSY search limits; however, the relic density constraint, among others, requires the lightest sneutrino mass to be more than $\sim 50$ GeV. Though this is not consistent with the CoGeNT-preferred range of $\sim 10$ 
GeV~\cite{cogent} for the DM mass, it is within the $2\sigma$-preferred range of CRESST-II~\cite{cresst} and also close to one of the DAMA-preferred mass 
range~\cite{dama}. Moreover, the benchmark points we find around 50 GeV sneutrino DM mass are all consistent with the recent hints of the lightest Higgs mass around 125 GeV~\cite{cms,atlas} which is very difficult 
to accommodate for a light neutralino DM in the usual cMSSM scenario~\cite{beskidt,baer2}.      

This paper is organized as follows: in Section 2, we briefly discuss the SUSY Inverse Seesaw Model (SISM) parameters and set up our notation; in Section 3, we present some benchmark points for the sneutrino DM in SISM satisfying all the existing experimental constraints; in Section 4, we discuss the collider signatures of a sneutrino LSP through cascade decays of squarks and gluinos and identify the same-sign dilepton+jets+large $\mET$~ signal; in Section 5, we present a detailed collider simulation of this signal for our benchmark points, along with the relevant SM background,  and propose some distinctions with the usual cMSSM case having similar squark-gluino spectrum; and in Section 6, we summarize our results.   
\section{The Model}
In order to explain the non-zero neutrino masses by an inverse seesaw mechanism~\cite{inverse}, the MSSM field content is supplemented by three pairs of SM-singlet superfields, (Dirac) $\hat{N}_i$ and (Majorana) $\hat{S}_i$ ($i=1,2,3$ for three generations)\footnote{Only one pair of $SU(2)_L$-singlets is sufficient to satisfy the neutrino oscillation data~\cite{hirsch}. However, if we want to generate all the neutrino masses at the tree-level, 
we must have three pairs of singlets.}. The superpotential is given by
\begin{eqnarray}
	{\cal W}_{\rm SISM}={\cal W}_{\rm MSSM}+\epsilon_{ab}y_\nu^{ij}\hat{L}^a_i\hat{N}_j\hat{H}^b_u + M_{R_{ij}}
\hat{N}_i\hat{S}_j+\mu_{S_{ij}}\hat{S}_i\hat{S}_j
\end{eqnarray}
where the $\mu_S$-term is the only lepton-number breaking term in the superpotential. The corresponding soft SUSY-breaking Lagrangian is
\begin{eqnarray}
	{\cal L}_{\rm SISM}^{\rm soft} &=& {\cal L}_{\rm MSSM}^{\rm soft} -\left[m_N^2\widetilde{N}^\dag\widetilde{N}+m_S^2\widetilde{S}^\dag\widetilde{S}\right]
-\left[\epsilon_{ab}A_\nu^{ij}\widetilde{L}^a_i\widetilde{N}_jH_u^b+B^{ij}_{M_R}\widetilde{N}_i\widetilde{S}_j+B_{\mu_S}^{ij}\widetilde{S}_i\widetilde{S}_j+{\rm h.c.}\right]\nonumber\\ 
\end{eqnarray}
The tree-level $9\times 9$ neutrino mass matrix in the basis $\{\nu_L,N^c,S\}$ is given by
\begin{eqnarray}
	{\cal M}_\nu = \left(\begin{array}{ccc}
		{\bf 0} & M_D & {\bf 0}\\
		M_D^T & {\bf 0} & M_R \\
		{\bf 0} & M_R^T & \mu_S
	\end{array}\right)
	\label{eq:mbig}
\end{eqnarray}
where $M_D=v_u y_\nu$ is the Dirac neutrino mass matrix, $v_u$ being the vacuum expectation value (vev) of the $\hat{H}_u$-superfield. The $3\times 3$ light neutrino mass matrix in the approximation $\mu_S\ll M_D<M_R$ is given by
\begin{eqnarray}
	M_\nu = \left(M_DM_R^{-1}\right)\mu_S\left(M_DM_R^{-1}\right)^T \equiv F\mu_S F^T
	\label{eq:vmass}
\end{eqnarray}
Assuming a TeV-scale inverse seesaw, $M_R\sim{\cal O}$(TeV), and ${\cal O}(0.1)$ Dirac Yukawa coupling (i.e., $M_D\sim {\cal O}(10)$ GeV), we need the lepton-number violating mass term $\mu_S\sim {\cal O}$(keV) for a sub-eV light neutrino mass, as required by the neutrino oscillation data~\cite{pdgvu}. Here we emphasize the fact that $\mu_S$ is much smaller than the other energy scale(s) pertinent to the SUSY sector. This feature, namely, the lepton number violation at a very low scale, is the quintessence of the inverse seesaw mechanism, which is integrated with the SUSY scheme here. The smallness of $\mu_S$ is 
technically natural in the 't Hooft sense, but must have its origin from some other new physics, e.g., radiative corrections~\cite{rad} or extra dimensions~\cite{ilmo}.

We note here that the gauge symmetry $SU(2)_L\times U(1)_Y$ 
allows for additional entries in the singlet sector, i.e., non-zero $\nu_LS$- and $NN$-terms in the neutrino mass matrix given by Eq.~(\ref{eq:mbig}). 
However, the presence of only the $NN$ term does not spoil the inverse seesaw structure at tree-level since the rank of mass matrix still remains the same.
But the $\nu_LS$ term will, in general, affect the inverse seesaw formula given by Eq.~(\ref{eq:vmass}), unless the coupling $y_S$ in the corresponding $y_S\hat{L}\hat{H}_u\hat{S}$-term in the superpotential is $\lsim 10^{-12}$ or so. 
These issues can be naturally eliminated by extending the SM gauge group so that these additional terms in the superpotential are forbidden by some symmetry (see, for instance, Refs.~\cite{khalil2,model}). However, in order to allow a direct comparison with the usual cMSSM case, we choose to work within the MSSM gauge group $SU(2)_L\times U(1)_Y$ and assume $y_S=0$. 

The mixing in the light neutrino sector is usually described by the unitary Pontecorvo-Maki-Nakagawa-Sakata (PMNS) matrix $U$ which diagonalizes the light neutrino mass matrix:
\begin{eqnarray}
U^TM_\nu U = {\rm diag}(m_1,m_2,m_3)
\label{eq:diaglight}
\end{eqnarray}
Since the above diagonalization of $M_\nu$ does not simultaneously diagonalize the 
other mass matrices $M_R$ and $\mu_S$ appearing in the full neutrino mass matrix given by Eq.~(\ref{eq:mbig}), there will be, in general, additional non-unitary contributions to the light neutrino mixing matrix due to its mixing with the heavy neutrinos. This can be derived from the $9\times 9$ unitary matrix ${\cal V}$ which diagonalizes the full neutrino mass matrix given by Eq.~(\ref{eq:mbig}): 
\begin{eqnarray}
{\cal V}{\cal M}_\nu {\cal V}^T = {\rm diag}(m_i,m_{R_j}),~ ~(i=1,2,3;~j=1,2,...,6)
\end{eqnarray}
by decomposing it into the blocks
\begin{eqnarray}
{\cal V}_{9\times 9}=\left(\begin{array}{cc}
{\cal U}_{3\times 3} & {\cal K}_{3\times 6}\\
{\cal K}^\dag_{6\times 3} & {\cal N}_{6\times 6}
\end{array}\right).
\label{eq:diagfull}
\end{eqnarray}
Then the upper $3\times 3$ sub-block will represent the full (non-unitary) light neutrino mixing matrix. 
To leading order in $F=M_DM_R^{-1}$, this can be expressed in terms of the PMNS matrix as follows:  
\begin{eqnarray}
{\cal U}\simeq \left({\bf 1} - \frac{1}{2}FF^\dag\right)U \equiv 
(1-\eta)U
\label{eq:nonunitary}
\end{eqnarray}
where $\eta=\frac{1}{2}FF^\dag$ measures the non-unitarity of the light neutrino mixing matrix.  

In the corresponding scalar sector, the sneutrino mass matrix is a $9\times 9$ complex, or $18\times 18$ real matrix which 
can be decomposed into two $9\times 9$ block-diagonals assuming 
$CP$-conservation in the corresponding soft-breaking sector:
\begin{eqnarray}
	{\cal M}_{\widetilde{\nu}}^2 &=& \left(\begin{array}{cc}
		{\cal M}_+^2 & {\bf 0} \\
		{\bf 0} & {\cal M}_-^2
	\end{array}\right)~ ~ {\rm with} \label{eq:svmass}
\\ 
	{\cal M}_\pm^2 &=& \left(\begin{array}{ccc}
		m_L^2+M_DM_D^\dag+\frac{1}{2}m_Z^2\cos 2\beta & \pm(v_uA_\nu-\mu M_D\cot\beta) & M_DM_R^\dag\\
		\pm(v_uA_\nu-\mu M_D\cot\beta) & m_N^2+M_RM_R^\dag+M_DM_D^\dag & \mu_S M_R^\dag\pm B_{M_R} \\
		M_DM_R^\dag & \mu_SM_R^\dag\pm B_{M_R} & m_S^2+\mu_S^2+M_RM_R^\dag\pm B_{\mu_S}
	\end{array}\right),\nonumber
\end{eqnarray}
where the corresponding mass eigenstates are linear combinations of the three sneutrino flavor eigenstates: $\widetilde{\nu}_{R_i,I_{j}}=\left(\widetilde{\nu}_{L_a},\widetilde{N}_b^c,\widetilde{S}_{d}\right)~(i,j=1,2,...,9;~a,b,d=1,2,3)$. In the next section, we examine the SUSY parameter space in which the lightest of these mass eigenstates can be the LSP. 
\section{Some Benchmark Points}
Our goal in this section is to find a sparticle spectrum with light sneutrino LSP in the cMSSM scenario with 
5 parameters ($m_0,m_{1/2}, \tan\beta,A_0,{\rm sign}~\mu$) and the additional 
inverse seesaw parameters $\mu_S,M_R,M_D,B_{\mu_S}$ and $B_{M_R}$. Once we find a light sneutrino LSP, we impose the relic density and 
direct detection constraints in order for it to qualify as a DM candidate. We also require all the benchmark points to satisfy various collider and low-energy constraints, summarized in Table~\ref{tab:expt}. 
\begin{table}[h!]
\begin{center}
\begin{tabular}{||c|c|c||}\hline\hline
Quantity & Value & Source \\ \hline\hline
$G_F$ & $1.1663787(6)\times 10^{-5}~{\rm GeV}^{-2}$ & \cite{pdg}\\
$\alpha_s(m_Z)$ & $0.1184\pm 0.0007$ & \cite{bethke}\\
$m_Z$ & 91.1876(21) GeV & \cite{pdg}\\
$m_\tau$ & 1.77682(16) GeV & \cite{pdg}\\
$m_b$  & $4.19\pm 0.12$ GeV & \cite{pdg}\\
$m_t$ & $173.2\pm 0.9$ GeV & \cite{tevmt}\\
$m_h$ & $125.3\pm 0.7$ GeV & \cite{cms}\\
$\Gamma_Z^{\rm invisible}$ & $<3.0$ MeV & \cite{lep} \\ \hline
$\Omega_{\rm CDM}h^2$ & $0.112\pm 0.006$ & \cite{wmap}\\
$\sigma_{\rm SI}$ & $<5\times 10^{-9}$ pb &\cite{xenon}\\ 
$\langle \sigma_A v\rangle$ & $< 10^{-26}~{\rm cm}^3{\rm s}^{-1}$ &\cite{fermilat}\\
\hline
$\Delta a_\mu$ & $(26.1\pm 8.0)\times 10^{-10}$ & \cite{amu}\\ 
$\Delta a_e$ & $(109\pm 83)\times 10^{-14}$ & \cite{ae}\\
\hline
BR($B\to X_s\gamma$) & $(3.21\pm 0.33)\times 10^{-4}$ & \cite{bsgamma}\\
BR($B_s^0\to \mu^+\mu^-$) & $<4.5\times 10^{-9}$ & \cite{bsmumu} \\ \hline
BR($\mu\to e\gamma$) & $<2.4\times 10^{-12}$ & \\
BR($\tau\to e\gamma$) & $<3.3\times 10^{-8}$ & \\
BR($\tau\to \mu\gamma$) & $<4.4\times 10^{-8}$ & \\
BR($\mu\to 3e$) & $<1.0\times 10^{-12}$ & \cite{pdg} \\
BR($\tau\to 3e$) & $<2.7\times 10^{-8}$ & \\
BR($\tau\to 3\mu$) & $<2.1\times 10^{-8}$ & \\ 
BR($\tau\to e\mu\mu$) & $<1.7\times 10^{-8}$ & \\
BR($\tau\to ee\mu$) & $<1.5\times 10^{-8}$ & \\
\hline
$|\eta|_{ee}$ & $0.002\pm 0.005$ & \\
$|\eta|_{\mu\mu}$ & $0.003\pm 0.005$ & \\
$|\eta|_{\tau\tau}$ & $0.003\pm 0.005$ & \\
$|\eta|_{e\mu}$ & $<7.2\times 10^{-5}$ & \cite{abadanuty}\\
$|\eta|_{e\tau}$ & $<1.6\times 10^{-2}$ & \\
$|\eta|_{\mu\tau}$ & $<1.3\times 10^{-2}$ & \\
\hline\hline
\end{tabular}
\end{center}
\caption{Various experimental constraints used in our analysis to find the benchmark points.}
\label{tab:expt}
\end{table} 
A few comments:
\begin{itemize}
	\item For the lightest Higgs mass, we use the CMS suggested value of $125.3\pm 0.4({\rm stat})\pm 0.5 ({\rm syst})$ GeV~\cite{cms}. The ATLAS suggested central value is around 126.5 GeV~\cite{atlas} with presumably similar experimental uncertainties; for concreteness, we just choose to work with the CMS value which has the errors explicitly stated. 
	\item For the spin-independent DM-nucleon scattering cross-section, we use the $2\sigma$ upper limit from the latest XENON100 data~\cite{xenon}. 
\item There also exist strong constraints on DM annihilation cross sections from indirect detection searches, e.g., in gamma rays~\cite{fermi,fermilat} and in high-energy neutrinos from the Sun~\cite{sato}. Here we use the latest 95\% confidence level upper limits obtained from the Fermi Large Area Telescope data~\cite{fermilat}.    
\item The lepton anomalous magnetic moments as shown in Table~\ref{tab:expt} are defined as $\Delta a_\ell=a_\ell^{\rm SM}-a_\ell^{\rm expt}$ where $a_\ell = (g-2)_\ell/2$. The most important one is the muon anomalous magnetic moment which persistently shows a $3\sigma$ discrepancy~\cite{amu} over the SM prediction and should be taken into account in any complete beyond SM scenario. For the electron $(g-2)$, the discrepancy is quite small and is a rather loose constraint on the new physics parameter space.  We do not consider the tau anomalous magnetic moments here, because its value is not known so precisely~\cite{pdg}. 
\item The non-unitarity of the light neutrino mixing matrix is defined in Eq.~(\ref{eq:nonunitary}) and the constraints on its elements shown in Table~\ref{tab:expt} are 
	derived from a combination of the neutrino oscillation data, the LEP precision data from weak gauge boson decays and the lepton-flavor violating (LFV) decays~\cite{abadanuty}.  
\end{itemize}

There are also strong constraints on the cMSSM parameter space from direct SUSY searches at the LHC~\cite{cmssusy,atlassusy}. Therefore, we must choose the input points in the $(m_0,m_{1/2})$-plane not already excluded by the LHC SUSY searches which for certain cases extend to $m_{1/2}\sim 600$ GeV and $m_0\sim 1$ TeV (e.g., in the jets+$\mET$~channel~\cite{jetmet}). On the other hand, very large values of $m_0$ and $m_{1/2}$ (larger than a few TeV) are not desirable from phenomenological perspective as they drive most of the sparticle masses beyond the kinematic reach of the LHC. 
Therefore, we choose our $m_0$ values close to 1 TeV and the $m_{1/2}$ values close to 600 GeV. We also choose to work with $\mu>0$ case, since $\mu<0$ is strongly disfavored by the muon anomalous magnetic moment as well as by the $B\to X_s\gamma$ branching ratio~\cite{mulz}. Similarly, large $\tan\beta$ values $\gsim 50$ are disfavored by the recent LHCb results on $B_s\to \mu^+\mu^-$~\cite{bsmumu}, and hence, we choose some intermediate values between 25 and 35 for the benchmark points discussed below. For the trilinear term $A_0$, the recent LHC discovery of a SM Higgs-like particle at 125 GeV~\cite{cms,atlas} implies that we must have a large negative $A$-term (for $\mu>0$) in order to have the radiative corrections account for the required enhancement of the lightest Higgs mass from its tree level value close to $m_Z$~\cite{martin}. 
 
In the neutrino sector, for simplicity, we assume the inverse seesaw parameter matrices $M_D,M_R$ as well as the $B$-terms $B_{\mu_S},B_{M_R}$ to be diagonal\footnote{We can choose this kind of texture since we are not working within any particular GUT framework.}. Hence, we can easily satisfy the LFV constraints for our benchmark points. Allowing non-zero off-diagonal entries in the Dirac Yukawa coupling matrix $y_\nu$ will induce large LFV effects, and we find that for the benchmark points discussed in the following section, we must have the off-diagonal entries less than $\sim {\cal O}(0.01)$ in order to satisfy all the LFV decay modes listed in Table~\ref{tab:expt}. Moreover, we assume no $CP$-violation in the neutrino sector, and choose all the mass matrices to be real\footnote{The addition of one or more $CP$-phases in the neutrino sector will not affect the sparticle spectrum, and hence, is irrelevant for our main results in the subsequent sections.}.    
Fixing both $M_D$ and 
$M_R$ also fixes the lepton-number breaking Majorana mass matrix $\mu_S$ 
by fitting to the neutrino mass and mixing parameters (assuming a particular mass hierarchy for the light neutrinos). 
Also note that since we are assuming a complete unification of the scalar sector, we choose $m_L^2=m_N^2=m_S^2=m_0^2$ at the high scale and similarly for the $A$-terms. 

The input parameters are chosen in such a way that all the experimental constraints listed in Table~\ref{tab:expt} are satisfied for all the benchmark points. Table~\ref{tab:input} lists all the input parameters for three benchmark points we have chosen to work with. For the low-energy values of $y_\nu$ and $M_R$ obtained by the renormalization group evolution of the parameters given in Table~\ref{tab:input}, the observed neutrino mass and mixing parameters can be fitted using appropriate values for the mass matrix $\mu_S$ in Eq.~(\ref{eq:vmass}). As an example, for a normal hierarchy of light neutrino masses, using the latest global fit values for the neutrino oscillation parameters~\cite{valleglobal} which includes the most recent $\theta_{13}$ results from Double CHOOZ, Daya Bay and RENO experiments:
\begin{eqnarray}
&& \Delta m^2_{21} = (7.62\pm 0.19)\times 10^{-5}~{\rm eV}^2,~ ~ \Delta m^2_{31} = (2.53\pm 0.09)\times 10^{-3}~{\rm eV}^2,\nonumber\\
&& \sin^2\theta_{12} = 0.320\pm 0.016,~ ~ \sin^2\theta_{23} = 0.490\pm 0.065,~ ~ \sin^2\theta_{13} = 0.026\pm 0.004,\nonumber
\end{eqnarray}  
we obtain the $\mu_S$ values as shown in Table~\ref{tab:input}. 
\begin{table}[h!]
\begin{center}
\begin{tabular}{||c|c|c|c||}\hline\hline
Input parameter & BP1 & BP2 & BP3 \\ 
\hline\hline
$m_0$ (GeV) & 993.68 & 996.84 & 815.79\\ 
$m_{1/2}$ (GeV)  & 600& 650 & 600 \\
$A_0$ (GeV) & $-2712.11$ & $-2858.42$ & $-2442.11$\\
$\tan\beta$ & 35 &25 & 30\\ \hline
$y_\nu$  & (0.16,0.16,0.18) & (0.10,0.10,0.08)& (0.10,0.10,0.10)\\
$M_R$ (GeV) & (300,1000,1000)& (200,1000,1000)& (610,1000,1000)\\
$B_{\mu_S}$ (GeV$^2$) & 10 & 10& 10\\
$B_{M_R}$ (GeV$^2$) & $10^6$ & $10^6$ & $10^6$\\
$\mu_S$ (eV) & \scriptsize $\left(\begin{array}{ccc}
2.04 & 8.27 & -5.34\\
8.27 & 56.69 & 12.31\\
-5.34 & 12.31 & 79.26\\
\end{array}\right)$ & 
\scriptsize $\left(\begin{array}{ccc}
2.25 & 12.82 & -9.69\\
12.82 & 123.64 & 31.39 \\
-9.69 & 31.39 & 236.39
\end{array}\right)$ & 
\scriptsize $\left(\begin{array}{ccc}
19.77 & 38.01 & -23.11\\
38.01 & 123.87 & 25.31\\
-23.11 & 25.31 & 153.34
\end{array}\right)$\\
\hline\hline
\end{tabular}
\end{center}
\caption{The input parameters for three chosen benchmark points (BP). The mSUGRA parameters are defined at the high scale and the singlet neutrino parameters at the low scale. We assume $\mu>0$ throughout and the neutrino sector parameters shown here have been chosen to be diagonal, except for $\mu_S$, as discussed in the text.}
\label{tab:input}
\end{table} 

\begin{table}[h!]
	\begin{center}
\small		\begin{tabular}{||c|c||c|c|c||}\hline\hline
Sparticle	&	Notation & BP1 & BP2 & BP3 \\ \hline
&	$(\widetilde{\nu}_{I_1},\widetilde{\nu}_{R_1})$ & (53.2155,53.3030) & 
(53.4623,53.5529) & (62.6587,62.7365)   \\  
&	$(\widetilde{\nu}_{I_2},\widetilde{\nu}_{R_2})$ & (834.7887,834.7890) & (953.3586,953.3598) & (743.3109,743.3119) \\  
&	$(\widetilde{\nu}_{I_3},\widetilde{\nu}_{R_3})$ & (930.6762,930.6810)  & (965.9735,965.9784) & (785.4476,785.4536) \\  
&	$(\widetilde{\nu}_{I_4},\widetilde{\nu}_{R_4})$ & (951.2057,951.2105) & (987.8791,987.8829) & (798.8994,798.9046) \\  
Sneutrino &	$(\widetilde{\nu}_{I_5},\widetilde{\nu}_{R_5})$ & (1033.8279,1033.8280) & (1065.9683,1065.9683) & (890.1739,890.1739)\\  
&	$(\widetilde{\nu}_{I_6},\widetilde{\nu}_{R_6})$ & (1042.0259,1042.0261) & (1068.5116,1068.5118)  & (893.2873,893.2875)  \\  
&	$(\widetilde{\nu}_{I_7},\widetilde{\nu}_{R_7})$ & (1419.8892,1419.8929) & (1415.7879,1415.7916) & (1420.8748,1420.8784) \\  
&	$(\widetilde{\nu}_{I_8},\widetilde{\nu}_{R_8})$ & (1715.9050,1715.9081)  & (1723.9674,1723.9704) & (1627.6817,1627.6848) \\  
&	$(\widetilde{\nu}_{I_9},\widetilde{\nu}_{R_9})$ & (1717.9193,1717.9224) & (1726.3187,1726.3217) & (1627.9388,1627.9419) \\ \hline
	 & $\widetilde{e}_1$ &1018.4 & 1025.3 & 846.1 \\
		 & $\widetilde{e}_2$ &1039.4 & 1069.0  & 893.6 \\
Slepton		 & $\widetilde{\mu}_1$ & 1016.6 & 1024.4 & 844.9 \\
		 & $\widetilde{\mu}_2$ & 1036.4 & 1068.6 & 893.2 \\
		 & $\widetilde{\tau}_1$ & 513.4 & 769.3 & 493.4  \\
		 & $\widetilde{\tau}_2$ &856.0 & 973.0 & 768.0 \\ \hline
		 & $\widetilde{u}_1$ &1535.0 &1607.5 & 1434.0 \\
		 & $\widetilde{u}_2$ & 1569.3 &1645.7 &  1471.2 \\
		 & $\widetilde{c}_1$ &1535.0 & 1607.5  & 1433.9  \\
		 & $\widetilde{c}_2$ &1569.1 & 1645.6 & 1471.0 \\
		 & $\widetilde{t}_1$ &634.2 & 625.0 & 613.8 \\
Squark		 & $\widetilde{t}_2$ &1151.6  & 1247.1 & 1125.3  \\ \cline{2-5}
		 & $\widetilde{d}_1$ &1531.6 & 1603.4 & 1430.2  \\
		 & $\widetilde{d}_2$ &1571.1 & 1647.4 & 1473.1  \\
		 & $\widetilde{s}_1$ &1531.5 & 1603.3 & 1430.1 \\
		 & $\widetilde{s}_2$ & 1570.9 & 1647.3 & 1473.0  \\
		 & $\widetilde{b}_1$ &1087.8 & 1194.3 & 1061.8  \\
		 & $\widetilde{b}_2$ & 1304.0 & 1460.0 & 1265.4 \\ \hline
		  Gluino & $\widetilde{g}$ &1401.4 &1505.3 & 1392.6 \\ \hline
		&	$\widetilde{\chi}^0_1$ &264.3 & 286.2 & 261.8\\
Neutralino		&	$\widetilde{\chi}^0_2$ & 499.2 & 539.8 & 495.2 \\
		&	$\widetilde{\chi}^0_3$ & $-1376.5$ & $-1464.4$ & $-1295.3$ \\
		&	$\widetilde{\chi}^0_4$ & 1379.5 & 1467.5 & 1298.7\\ \hline
		Chargino & $\widetilde{\chi}^\pm_1$ & 499.4 & 540.0 & 495.4 \\
		 & $\widetilde{\chi}^\pm_2$ & 1380.1 & 1467.9 & 1299.2 \\ \hline
			\hline
	\end{tabular}
\end{center}
\caption{The sparticle masses (in GeV) for the chosen benchmark points. The sneutrino masses are shown up to four decimal places to illustrate the lifting of degeneracy between the mass eigenstate pairs due to the small lepton-number breaking.} 
\label{tab:spectra}
\end{table}

The low-energy mass spectrum for the superpartners corresponding to the three benchmark points are tabulated in Table~\ref{tab:spectra}.
For this purpose, we have used {\tt SARAH}~\cite{sarah} to implement the SISM scenario, and {\tt SPheno}~\cite{spheno} to generate the mass spectra and to evaluate some of the low-energy observables. The DM relic density and its   scattering and annihilation cross sections were calculated using {\tt micrOMEGAS}~\cite{micromegas}.

Note that the sneutrino real scalar fields 
$(\widetilde{\nu}_{I_i},\widetilde{\nu}_{R_i})$ are split in their masses with the mass splitting in the range of keV-MeV within each pair which is a characteristic feature of the SUSY inverse seesaw mechanism~\cite{adcm}. In later sections, we will sometimes denote the lightest mass eigenstate pair $(\widetilde{\nu}_{I_1},\widetilde{\nu}_{R_1})$ simply by $\widetilde{\nu}_1$.     

It is clear from Table~\ref{tab:spectra} that all the benchmark points satisfy the LHC direct search limits on the SUSY particle masses in cMSSM. They also satisfy the other low-energy experimental constraints in Table~\ref{tab:expt}, as shown in Table~\ref{tab:low}. Here we want to make some comments on these observables:
\begin{table}[h!]
\begin{center}
\begin{tabular}{||c|c|c|c||} \hline\hline
Parameter & BP1 & BP2 & BP3 \\ \hline\hline
$m_h$ (GeV) & 123.9 & 123.8 & 123.7 \\ \hline
$\Omega_{\rm DM}h^2$ &  0.105 & 0.106 & 0.119 \\ 
$\sigma_{\rm SI}$ (pb) & $8.2\times 10^{-9}$ & $1.3\times 10^{-9}$ & $2.4\times 10^{-9}$ \\ 
$\langle \sigma_A v\rangle~({\rm cm}^3{\rm s}^{-1})$ & $5.6\times 10^{-34}$ & $1.5\times 10^{-34}$ & $5.4\times 10^{-35}$ \\ 
\hline
$\Delta a_\mu$ & $5.1\times 10^{-10}$ & $3.3\times 10^{-10}$ & $5.4\times 10^{-10}$ \\
$\Delta a_e$ & $1.2\times 10^{-14}$ & $7.7\times 10^{-15}$ & $1.3\times 10^{-14}$ \\ \hline
BR$(B\to X_s\gamma)$ & $2.6\times 10^{-4}$ & $2.7\times 10^{-4}$ & $2.6\times 10^{-4}$ \\
BR$(B_s\to \mu^+\mu^-)$ & $3.6\times 10^{-9}$ & $3.7\times 10^{-9}$ & $3.7\times 10^{-9}$ \\ \hline
BR$(\mu\to e\gamma)$ & $8.0\times 10^{-21}$ & $8.2\times 10^{-22}$ & $4.4\times 10^{-21}$ \\
BR$(\tau\to e\gamma)$ & $1.0\times 10^{-19}$ & $1.6\times 10^{-20}$ & $8.1\times 10^{-20}$\\
BR$(\tau\to \mu\gamma)$ & $1.4\times 10^{-15}$ & $2.4\times 10^{-16}$ & $1.3\times 10^{-15}$ \\
BR$(\mu\to 3e)$ & $6.7\times 10^{-21}$ & $2.2\times 10^{-21}$ & $6.2\times 10^{-23}$\\
BR$(\tau\to 3e)$ & $1.7\times 10^{-19}$ & $4.5\times 10^{-20}$ & $1.1\times 10^{-21}$\\
BR$(\tau\to 3\mu)$ & $3.6\times 10^{-15}$ & $6.2\times 10^{-16}$ & $4.8\times 10^{-17}$\\ \hline
$|\eta_{ee}|$ & $4.3\times 10^{-3}$ & $3.9\times 10^{-3}$ & $4.5\times 10^{-4}$\\
$|\eta_{\mu\mu}|$ & $3.4\times 10^{-4}$ & $1.6\times 10^{-4}$ & $1.6\times 10^{-4}$\\
$|\eta_{\tau\tau}|$ & $2.9\times 10^{-4}$ & $9.8\times 10^{-5}$ & $1.5\times 10^{-4}$\\ \hline\hline
\end{tabular}
\end{center}
\caption{The low-energy observables for the three chosen BPs. These values are to be compared with the experimental values in Table 1.}
\label{tab:low}
\end{table}
\begin{itemize}
\item It is well known that a 125 GeV mass for the lightest neutral Higgs boson in MSSM is not very natural~\cite{125gev}. It becomes even more difficult in cMSSM if one has to satisfy the other low-energy constraints and requires the neutralino LSP to have the observed relic density~\cite{beskidt,baer2}. The situation is somewhat similar in our case; however, since fixing the exact mass of the suspected scalar resonance at the LHC will require more data, we are content with values within 2 GeV of the average of the CMS and ATLAS central values. We believe that any tweaking of parameters to confirm the exact Higgs mass, when it is known with greater precision, will not affect the general conclusions of this paper.
\item The correct relic density is obtained near the resonant enhancement region of the annihilation cross-section in the Higgs-mediated $s$-channel process: $\widetilde{\nu}_1 \widetilde{\nu}_1\to f\bar{f}$ where $f$ denotes the SM fermion (mostly $b$ and $\tau$ in our case). Therefore, all our benchmark points have the LSP mass close to $m_h/2$. This is illustrated in Figure~\ref{fig:relic} where we have plotted the relic density versus the  sneutrino LSP mass for some typical Dirac Yukawa coupling values. 
\begin{figure}[h!]
\centering
\includegraphics[width=7cm]{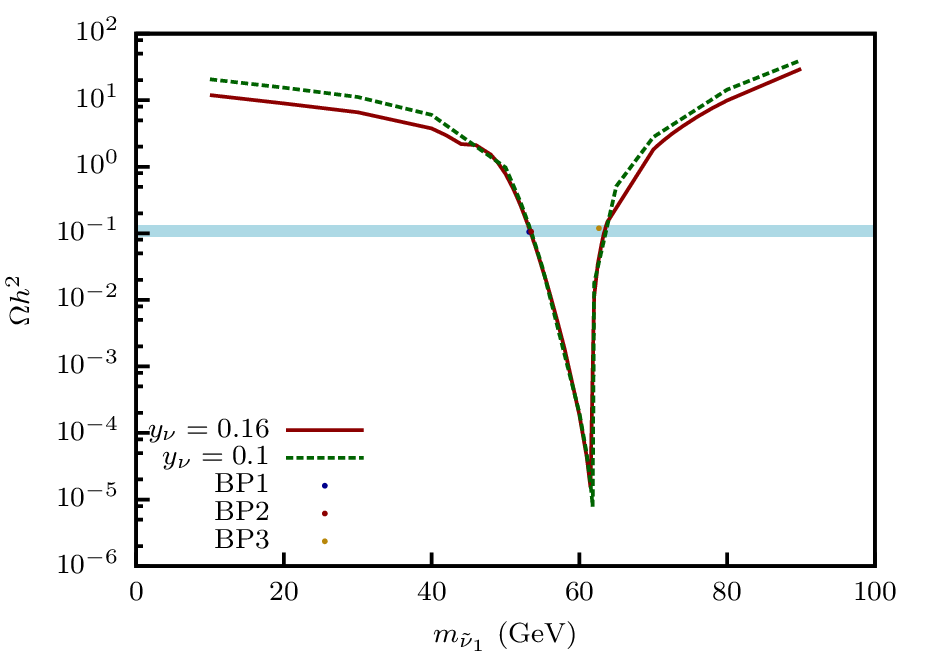}
\caption{The relic density as a function of the sneutrino LSP mass for the first generation Dirac Yukawa coupling values of 0.10 and 0.16 while keeping all other input parameters fixed. The points corresponding to our benchmark points are shown by dots. The shaded region is the $3\sigma$-range allowed by the WMAP data.}
\label{fig:relic}
\end{figure}
We also find that the sneutrino LSP-nucleon spin-independent elastic\footnote{Note that the splitting between the lightest sneutrino mass eigenstates for our benchmark points is of order of a few MeVs which is too large for inelastic DM-nucleon scattering to occur. Hence we only consider the elastic scattering.} cross section is within the $1\sigma$ upper bound of the latest XENON100 result except for BP1 which is slightly above their $2\sigma$ limit.   

\item The dominant annihilation channels for the sneutrino DM in our case have $b\bar b$ and $\tau\bar\tau$ final states. For small admixtures of the left sneutrino component (allowed by the $Z$-invisible decay width) in a mostly dominant singlet sneutrino LSP eigenstate, the thermally averaged annihilation cross section values for our benchmark points are well within the experimental upper limits given in Table~\ref{tab:expt}. 
  
\item The SUSY contributions to $\Delta a_\mu$ have been calculated to the two-loop order~\cite{susyamu} and it has been shown that for a range of parameter space, it is possible to accommodate the discrepancy. As can be seen from Tables~\ref{tab:expt} and \ref{tab:low}, we are able to explain the discrepancy 
	within $3\sigma$ for all our benchmark points. Also for the electron anomalous magnetic moment, we are consistent with the smaller discrepancy to 
	within $1\sigma$. 
\item The SM prediction for the branching ratio of the weak radiative $B$-meson decay is $(3.15\pm 0.23)\times 10^{-4}$~\cite{smbsgamma}, and comparing 
	with the most recent experimental value from the BaBar experiment as given in Table~\ref{tab:expt}, we see that there is very little room left for the SUSY 
	contribution~\cite{susybsgamma}. However, for the choice of our mSUGRA parameters, the SUSY contributions are negligible and the values predicted for all our BPs are  
	within $2\sigma$ of the experimental value. On the other hand, for the branching ratio of the flavor-changing-neutral-current (FCNC) process $B_s^0\to \mu^+\mu^-$, the SM 
	contribution is small: $(3.2\pm 0.2)\times 10^{-9}$~\cite{buras}; hence, comparable SUSY contributions~\cite{susybsmumu} are still allowed by the latest data from the LHCb experiment. We estimated that the branching ratios for our benchmark points are within this allowed range.     
\item  For the rare LFV decays, since we are working within an mSUGRA scenario, the SUSY contributions are quite small~\cite{gabbi}. But the contributions from the leptonic sector could be large in seesaw models with large Yukawas~\cite{lfvinverse}, as in our case. However, due to our choice of the diagonal textures for the Dirac Yukawas, the leptonic contributions also vanish altogether. Hence, we have very small LFV branching ratios for all the benchmark points.  
\item The non-unitarity effects could also, in principle, be large in low-scale inverse seesaw models with large Dirac Yukawas~\cite{model,hirsch,nutyinv}. In our case, again due to the diagonal textures chosen for both $M_D$ and $M_R$, the non-unitarity parameter, defined by Eq.~(\ref{eq:nonunitary}), is also a 
	diagonal matrix. Hence we only show the values for its diagonal entries in Table~\ref{tab:low}, and all our values are within the current experimental bounds.
\end{itemize}

We also note that in our scenario, since the sneutrino LSP is sufficiently light, the lightest neutral Higgs boson can, in principle,  decay into a pair of LSP's, thus giving rise to an invisible decay width of the Higgs boson. The LHC signatures of these decays are relatively clean, and very large branching ratios to an invisible decay channel are disfavored by the current LHC Higgs searches~\cite{ghosh}. The branching ratio depends, among other things, on the neutrino Yukawa coupling $y_\nu$. Recent global analyses~\cite{trott}\footnote{For a similar analysis with the earlier LEP/Tevatron/XENON/WMAP data and a 50-60 GeV scalar DM scenario (as in our case), see Ref.~\cite{yann}.} have reported that the present LHC Higgs data can indeed accommodate an invisible branching ratio for the Higgs boson, although their best fit 
values for this do not quite agree with each other. If such a possibility is more precisely fixed by future data, it may lead to an estimate of the bounds 
on the neutrino Yukawa couplings in the inverse seesaw models which could be compared with those obtained from direct Higgs search results~\cite{roberto}.

We have thus demonstrated convincingly that (a) a hybrid scenario for the origin of soft SUSY-breaking masses can be used consistently with the inverse seesaw mechanism, (b) one can have a sneutrino LSP which is light and is still consistent with all the existing experimental constraints, and (c) the rest of the SUSY spectrum is phenomenologically viable. The next question to ask is whether there are any distinctive signatures of this scenario which can be seen at the LHC. We address this question in the next section.  
\section{Collider Signatures}
\begin{figure}[h!]
	\begin{center}
\includegraphics[width=5cm]{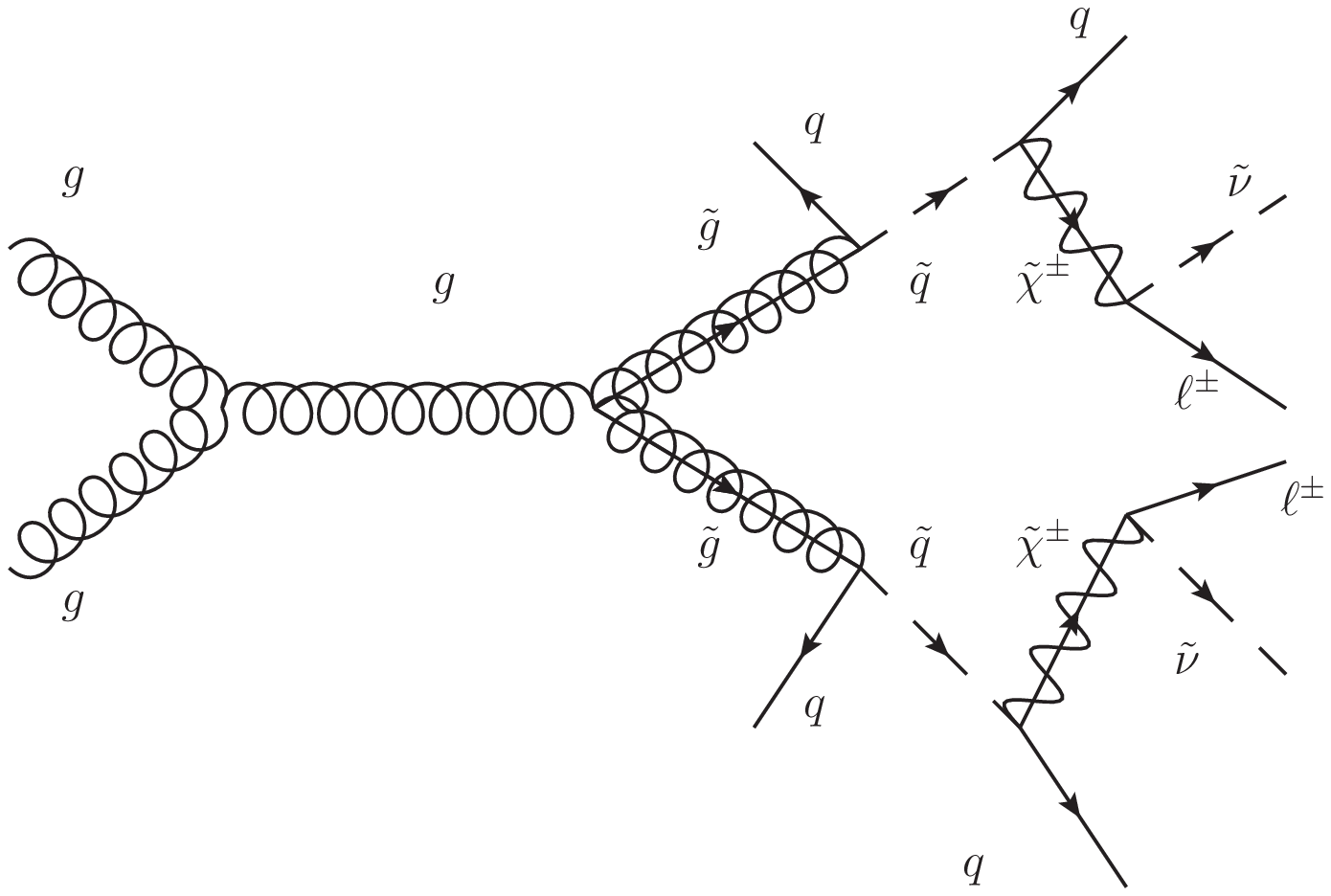}
\hspace{2cm}
\includegraphics[width=5cm]{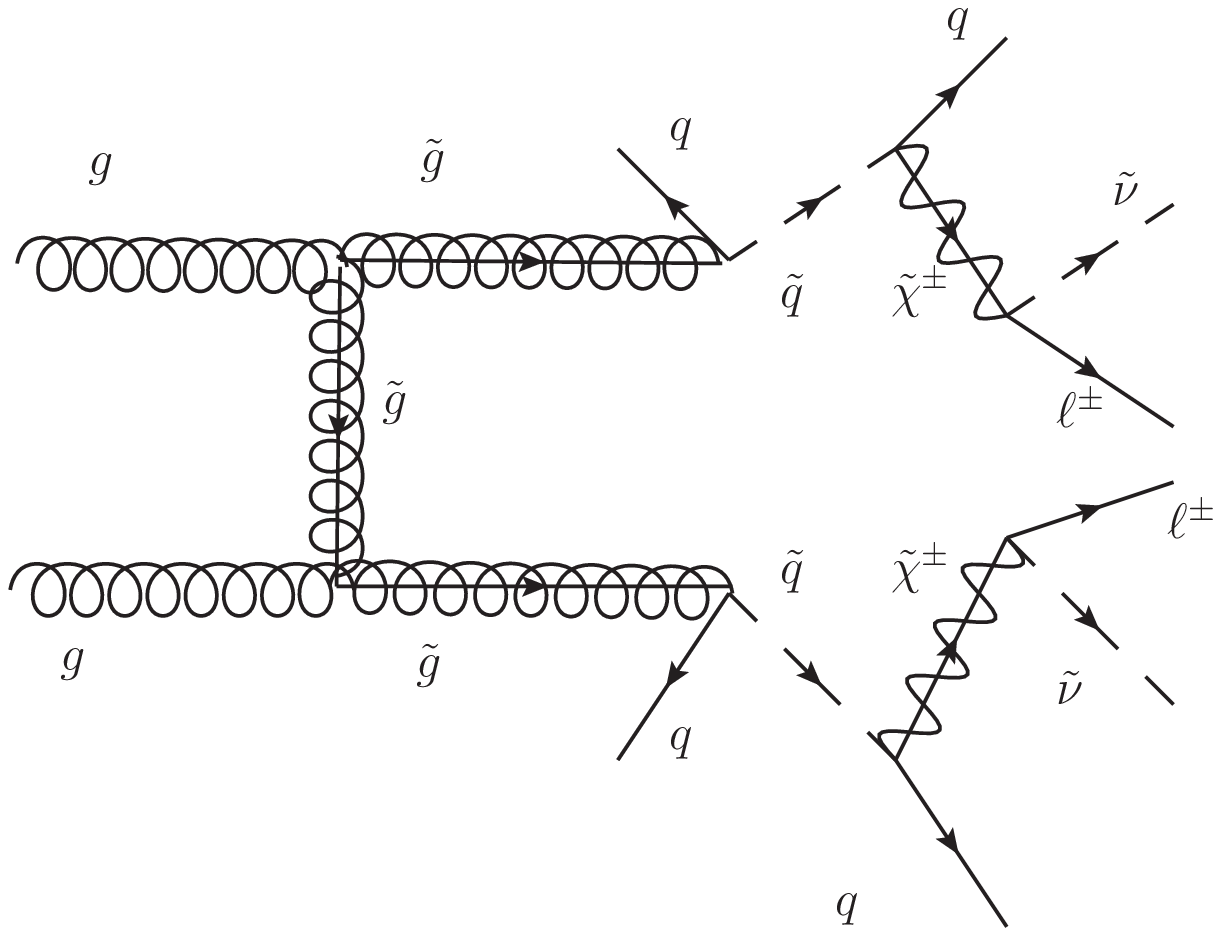}\\
\includegraphics[width=5cm]{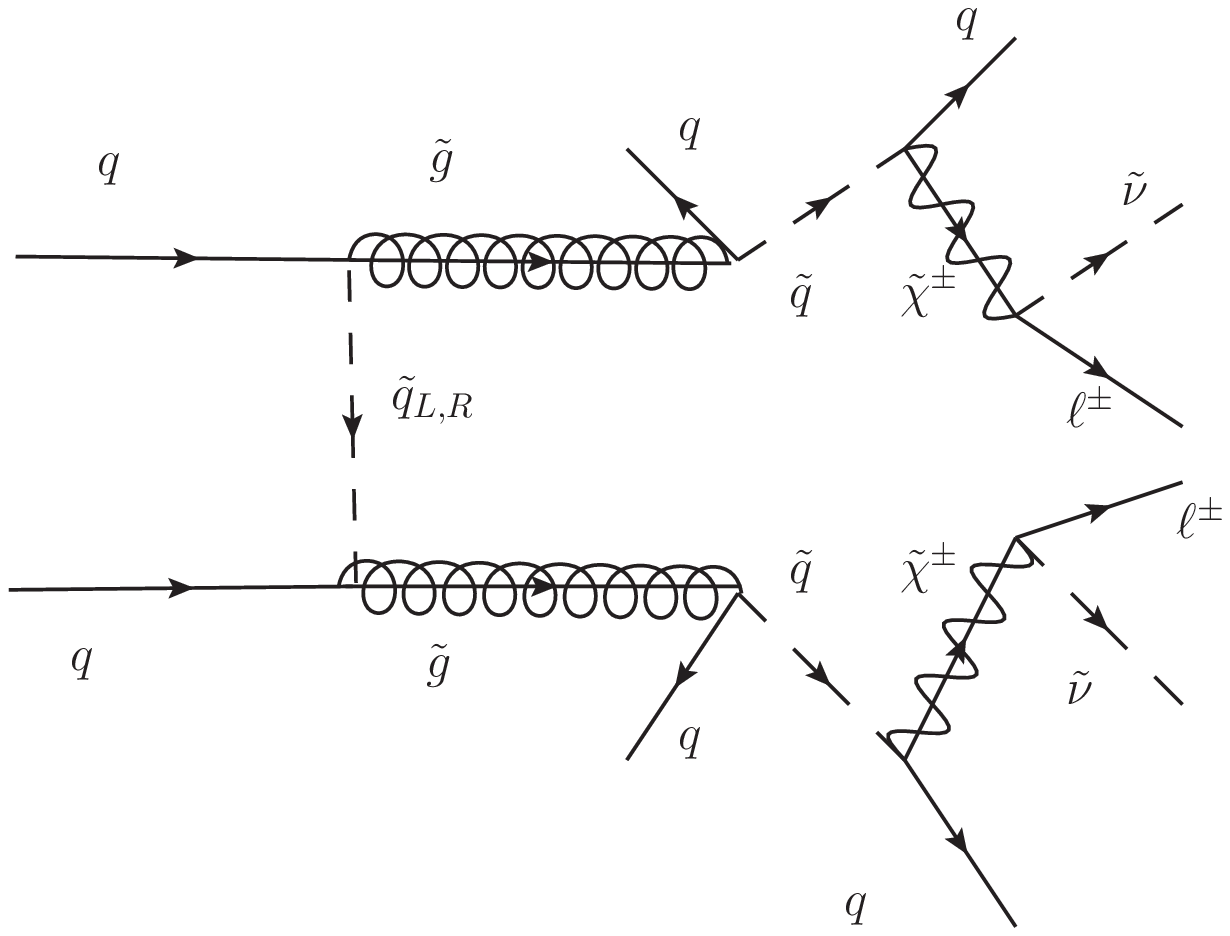}
\hspace{2cm}
\includegraphics[width=5cm]{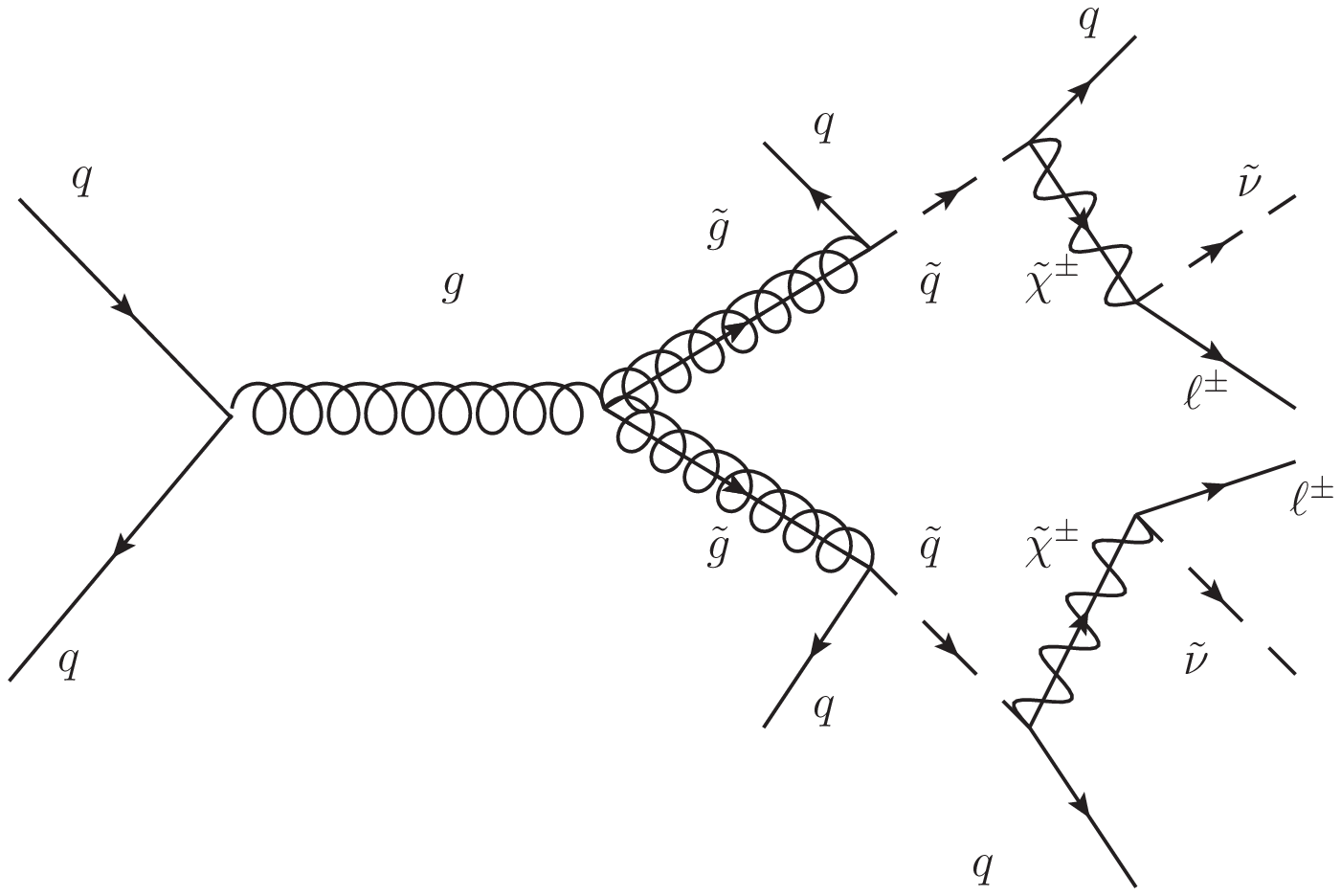}
\end{center}
\caption{The Feynman diagrams for the gluino pair-production and its cascade decays to give the same-sign dilepton+jets+$\mET$ signal at the LHC. Similar 
diagrams exist for squark-gluino and squark-squark production and decay which we have not shown here.} 
\label{fig:feyn1}
\end{figure}

The most copious collider signals of any SUSY scenario will come from the production of colored superpartners, namely, squarks and gluinos, which will have cascade decays through charginos and neutralinos, eventually ending up in the stable LSP in $R$-parity conserving SUSY models~\cite{baer}. Unless these squarks and gluinos are too heavy to be kinematically accessible, they will have substantial production cross sections at a hadron collider due to the strong interaction. The production channels are gluino-pair production, squark-gluino associated production and squark-squark pair production (see Fig.~\ref{fig:feyn1}). As the direct decay of the squarks and gluinos to the color- and electrically-neutral LSP are either forbidden or occur with only a tiny branching fraction, the dominant decay modes for the gluino always involve quarks (and hence multiple jets in the final states). The gluino can have either the two-body decay via $\widetilde{g}\to q\widetilde{q}$, if kinematically allowed, or the three-body decay modes $\widetilde{g}\to q\bar{q}'\widetilde{\chi}^\pm_i,~q\bar{q}\widetilde{\chi}^0_j$ with virtual squarks.  Similarly, the squarks decay to two-body modes $\widetilde{q}\to q\widetilde{g}$, if kinematically allowed, or $\widetilde{q}_L\to q'\widetilde{\chi}^\pm_i,~q\widetilde{\chi}^0_j$, while $\widetilde{q}_R\to q\widetilde{\chi}^0_j$ only, since right-handed squarks do not couple to charginos in the MSSM. 
If the squarks are degenerate, and the Yukawa coupling effects
negligible, the three-body decays to the wino-like charginos
and neutralinos usually have larger branching fractions due to their larger gauge couplings. If $|\mu|< M_2$, gluinos and squarks may thus decay most of the time to the heavier charginos and neutralinos, resulting in lengthier cascade decay chains than those shown in Figure~\ref{fig:feyn1}.

The LHC signals of our scenario can differ considerably from those of the usual cMSSM situation. To understand this, let us look at the branching ratios of the two-body decays of the lighter chargino ($\widetilde{\chi}_1^+$) which are listed in Table~\ref{tab:cha} for all of our benchmark points. While the conventionally expected decay $\widetilde{\chi}_1^+\to W^+\widetilde{\chi}_1^0$ is there, it is dependent on the $\widetilde{W}_3^0$ components of $\widetilde{\chi}_1^0$ as well as the Higgsino components of both $\widetilde{\chi}_1^+$ and $\widetilde{\chi}_1^0$. On the other hand, here we have another often dominant channel, namely, $\widetilde{\chi}_1^+\to \ell^+\widetilde{\nu}_1$ (where $\widetilde{\nu}_1$ is the sneutrino LSP and $\ell=e,\mu$) triggered by the large mixing in the sneutrino sector\footnote{$\ell$ is mostly electron for our choice of benchmark points, though cases with muons do not make any difference in our analysis.  
Since the tau-lepton detection efficiency is 
not as good as for the electron and muon, we will not analyze the tau-lepton final states.}. Consequently, the leptonic branching ratio of the $\widetilde{\chi}_1^+$ 
is remarkably enhanced\footnote{Note that all charged sleptons are heavier than the lighter chargino in our case.}. Thus the SUSY cascades lead to a highly boosted rate of dileptons, of which the same-sign dileptons (SSD) are more spectacular being relatively background-free. The scenario outlined by us will therefore exhibit a rise in the SSD rate with respect to that of purely jets+$\mET$ events, as compared to a cMSSM spectrum of comparable heaviness. We also expect the 
$\mET$ distribution to be different for a sneutrino LSP case than the neutralino LSP case, as noted earlier in Ref.~\cite{lessa}. In particular, the $\mET$-distribution is expected to be much harder in our SISM scenario compared to the cMSSM scenario. 
\begin{table}[h!]
	\begin{center}
		\begin{tabular}{||c|c|c|c||}\hline\hline
			Chargino ($\widetilde{\chi}_1^+$) decay & BP1 & BP2 & BP3\\ \hline\hline 
			 $W^+\widetilde{\chi}_1^0$ & 0.23 & 0.45 & 0.31\\ 
			 $\ell^+\widetilde{\nu}_1$ & 0.77 & 0.55 & 0.69\\
			\hline\hline
		\end{tabular}
	\end{center}
	\caption{The lighter chargino decay branching ratios for our benchmark points in SISM. On the other hand, in the cMSSM case with neutralino LSP, 
	the branching ratio is close to 100\% for the decay $W^+\widetilde{\chi}_1^0$. }
	\label{tab:cha}
\end{table}

To illustrate this SSD-enhancement effect in our case compared to the pure cMSSM scenario, we construct a ratio as follows: 
\begin{eqnarray}
r = \frac{\sigma(\ell^\pm\ell^\pm+\geq 2j+\mET)}{\sigma(0\ell+\geq 3j+\mET)}
\label{eq:r}
\end{eqnarray}
which is expected to be larger in our case, and as shown in the next section, could be used to distinguish our SISM scenario with sneutrino LSP from the usual cMSSM scenario with neutralino LSP.

Here we want to emphasize that the SSD signal in inverse seesaw is purely supersymmetric in nature. In other words, if one leaves aside the SUSY processes,  the SSD signal is suppressed due to the small lepton-number violation and pseudo-Dirac nature of the singlet neutrinos. In that case, however, one can 
look for the tri-lepton signals with $\mET$ for its LHC discovery potential~\cite{del,trilepton}. Note that one can also investigate the SUSY inverse seesaw in the tri-lepton channel in which case novel correlations of the tri-lepton signal with the neutrino mixing angles can be searched for~\cite{mondal}.    
\section{Event Generation, Background Simulation and Results}
In this section, we give a detailed description of the SSD+jets+$\mET$~ signal in our SISM case with light sneutrino LSP and a comparison of the signal strength with a canonical cMSSM scenario with neutralino LSP having similar squark-gluino spectrum for a possible distinction of the two cases at the $\sqrt s=14$ TeV LHC.
The SUSY spectrum and the various decay branching fractions were calculated using {\tt SPheno}~\cite{spheno}. The {\tt SLHA} file is then fed to {\tt{PYTHIA}} (version 6.409)~\cite{pythia} for event generation. The initial and final state quark and gluon radiation, multiple interactions, decay, hadronization, 
fragmentation and jet formation are implemented following the standard procedures in {\tt{PYTHIA}}. The factorization and renormalization scales are set at $\sqrt{\widehat s}~$
(i.e $\mu_R =\mu_F =\sqrt{\widehat s}$~), where $\sqrt{\widehat s}~$ is the parton level center of mass energy. 
We have used the leading order {\tt CTEQ5L} parton distribution 
functions~\cite{cteq} for the colliding protons.  The jets are constructed using the cone algorithm in {\tt PYCELL}; only those jets are constructed which have $p_T > 20~{\rm GeV}$ and $|\eta| < 2.5$. 
To simulate the detector effects, we have taken into account the smearing of jet energies by a Gaussian 
probability density function of width $\sigma (E)/E_j = (0.6/\sqrt{E_j[{\rm GeV}]}) + 0.03$ where $E_j$ is the unsmeared jet energy~\cite{Barr}.  

In order to find the same-sign di-leptons+$n$ jets+$\mET$ (with $n\geq 2$) final states, we impose the following selection criteria:
\begin{itemize}
\item $p^\ell_T>10 ~{\rm{GeV}}$ and $|\eta^\ell|<2.4$ for both the leptons. For the same-flavor dilepton final states, we raise it to $p_T^\ell>15$ GeV. 
\item Lepton-lepton separation $\Delta R_{\ell\ell}> 0.2$, where $\Delta R=\sqrt{(\Delta\eta)^2 + (\Delta\phi)^2}$.
\item Lepton-jet separation $\Delta R_{\ell j}> 0.4$.
\item The sum of $E_T$ deposits of the hadrons which fall within a cone of $\Delta R\le 0.2$ around a lepton, must be less than $0.2~p_T^\ell$. 
\item Jet-jet separation $\Delta R_{jj}> 0.4$.
\end{itemize}

\begin{figure}[h!]
\centering
\includegraphics[width=6cm]{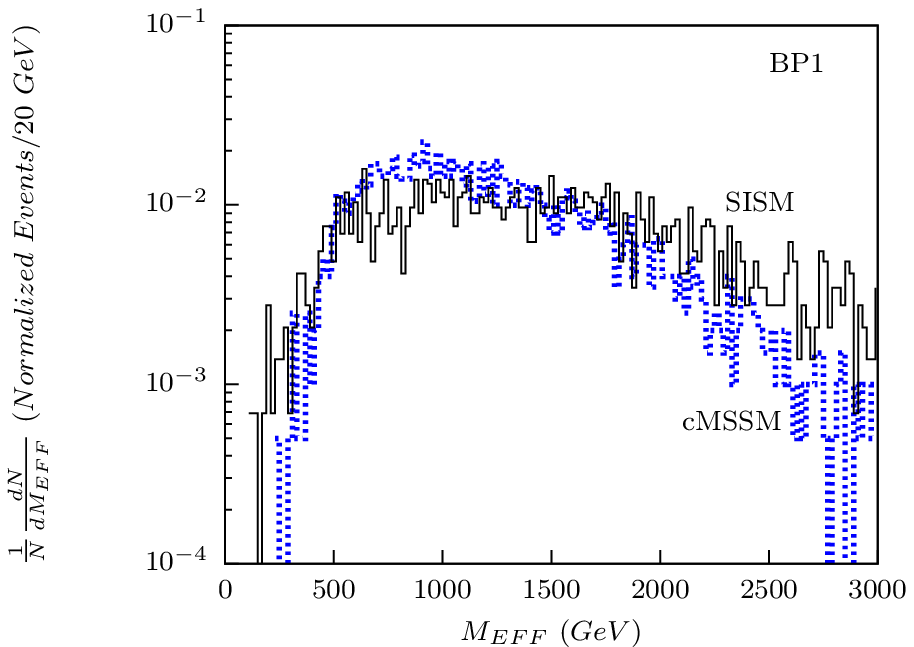}
\hspace{1cm}
\includegraphics[width=6cm]{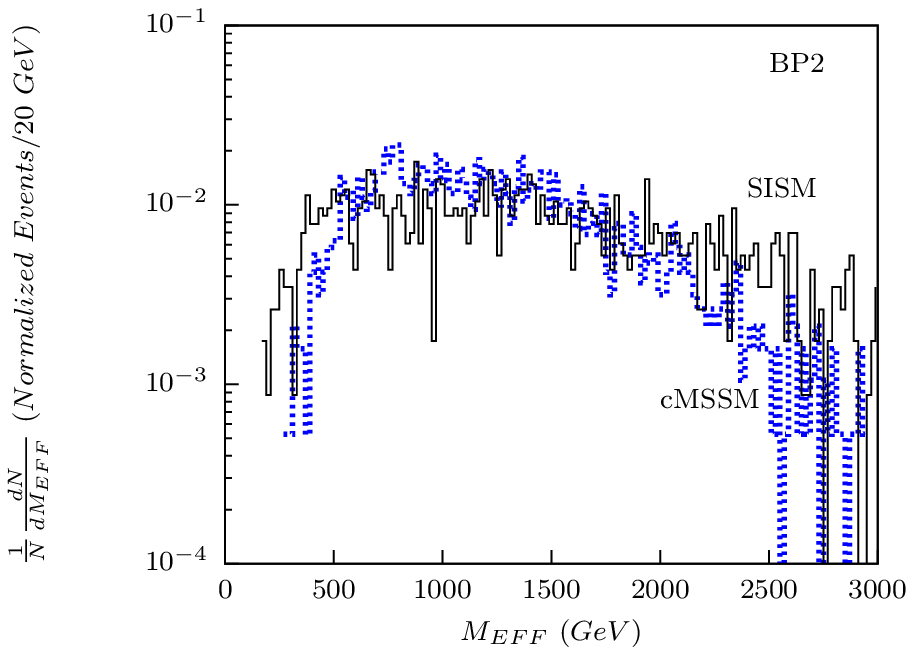}
\hspace{1cm}
\includegraphics[width=6cm]{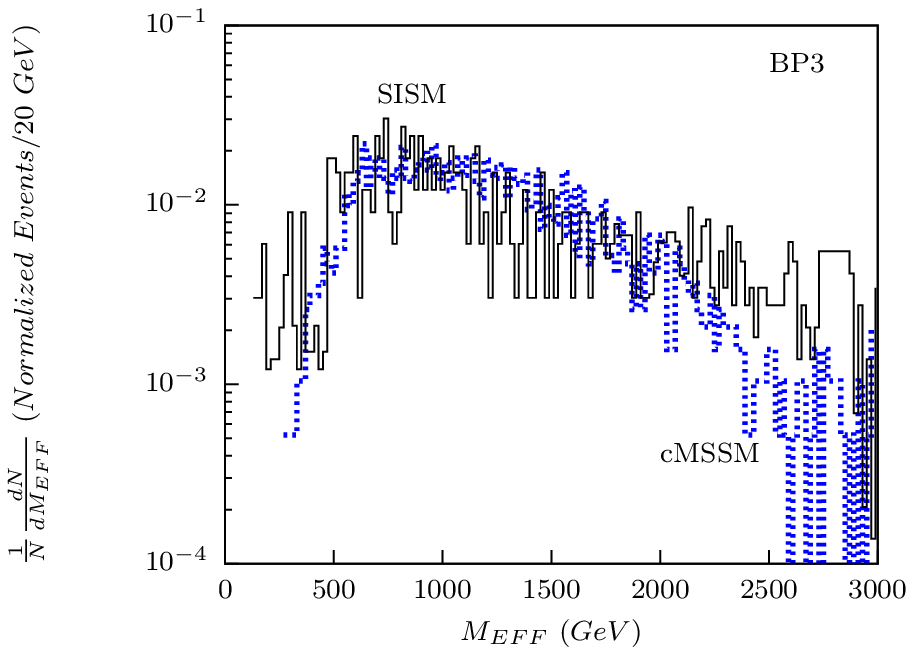}
\caption{The effective mass distribution of the final states as defined in Eq.~(
5.1) for the SISM and cMSSM scenarios to illustrate that the squark-gluino spectrum considered in both cases are similar.}
\label{fig:meff}
\end{figure}

Since our goal is to distinguish the SUSY inverse seesaw scenario from the conventional cMSSM case, we need to consider similar squark-gluino spectrum for both the cases. In order to do so, we generated similar benchmark points for the cMSSM case using the same mSUGRA input parameters given in Table~\ref{tab:input} and also checked that the effective mass distributions, defined as the scalar sums of the lepton and jet transverse momentum and missing transverse energy:
\begin{eqnarray}
M_{\rm eff} =  \sum |p_T^\ell|+\sum|p_T^j|+\mET~,
\label{eq:meff}
\end{eqnarray}
are similar for both the scenarios, as shown in Fig.~\ref{fig:meff} for all the benchmark points.

Now in order to distinguish the two scenarios, we compute the ratio $r$ defined in Eq.~(\ref{eq:r}) for both SISM and cMSSM cases which are tabulated in Table~\ref{tab:r}. We find that the value of $r$ in the SISM case is roughly $4-5$ times higher than the cMSSM case  for all the benchmark points. 
\begin{table}[h!]
	\begin{center}
		\begin{tabular}{||c|c|c|c||}\hline\hline
$r$ & BP1 & BP2 & BP3 \\ \hline
SISM &0.19 &0.15 &0.11 \\
cMSSM &0.04 &0.03 &0.03 \\ 
\hline\hline
\end{tabular}
\end{center}
\caption{The $r$ values for all the benchmark points in both SISM and cMSSM cases.} 
\label{tab:r}
\end{table}
Apart from this clear distinction, we also expect more $\mET$~ in the chargino decay in case of the SISM, as mentioned earlier. In order to illustrate this, we need to analyze the $\mET$ distribution for the SSD+$\geq 2j+\mET$~ signal for both SISM and cMSSM cases. We also need to analyze the SM backgrounds in detail, as follows.  

The dominant SM background for the SSD events come from $t\bar t,Wt\bar t,WWW,Wb\bar b,Zb\bar b,\\WWnj,WZnj,
ZZ$ final states at the LHC~\cite{del}. All the SM backgrounds except the $t\bar t$ were generated at the parton level using {\tt{ALPGEN}} (version 2.14) \cite{alpgen} with default 
factorization and renormalization scales, and then fed to  {\tt{PYTHIA}} for showering, hadronization, fragmentation, decay, etc.  The $t\bar t$ background was directly generated and analyzed in {\tt PYTHIA}. The number of events obtained after the selection criteria for $\sqrt s=14$ TeV LHC and normalized to 
30 fb$^{-1}$ luminosity are shown in Table~\ref{tab:signal}. Note that at this stage, some of the SM backgrounds are much larger than the SSD signal, and we need to devise further cuts to reduce the background without affecting the signal much. As shown in Table~\ref{tab:signal}, we found two relevant cuts, namely,  (a) $p_T^j>50$ GeV for all jets and $p_T^j>100$ GeV for the leading jet, and (b) $\mET>300$ GeV which reduce the SM background significantly. 
\begin{table}[h!]
\begin{center}
\begin{tabular}{| c|  c  c  c | c  c  c |c c c|} \hline 
 Channel &\multicolumn{3}{|c|}{After basic selection criteria} & \multicolumn{3}{|c|}{After jet-$p_T$ cut} & \multicolumn{3}{|c|}{After $\mET$ cut} \\ \cline{2-10}
 &$\mu\mu$  & $e\mu$  & $e e$  & $\mu\mu$  & $e\mu$  & $e e$ & $\mu\mu$  & $e\mu$  & $e e$ \\ \hline
BP1 & 33.24  & 125.18  & 144.01 & 30.73  & 112.66  & 127.45 & 24.30  & 90.13  & 114.69 \\
BP2 & 39.95  & 32.44   & 97.26  & 34.38  & 26.86   & 84.34 & 28.54  & 23.35  & 64.87 \\
BP3 & 35.94 & 88.94 & 102.84 & 34.15 & 80.05 & 91.49  & 32.44  & 78.48  & 86.76 \\  
\hline
$WWW$ & 16.86  & 12.36  & 29.64 & 3.18  & 2.49  & 6.00 & 0.39  & 0.24  & 0.24\\
$WWjj$ & 140.01 & 75.39  & 193.86 & 75.39  & 43.08  & 96.93 & 0.00  & 0.00  & 0.00\\
$WZ$  & 84.60  & 16.92  & 186.06 & 33.84  & 0.00  & 51.00 & 0.00  & 0.00  & 0.00\\
$ZZ$  & 0.33  & 0.33  & 0.66 & 0.000  & 0.000  & 0.03 & 0.00  & 0.00  & 0.00\\
$Wb\bar b$ & 29.25 & 5.85 & 29.25 & 0.00 & 0.00 & 0.00 & 0.00 & 0.00 & 0.00 \\
$Wt\bar t$ & 81.33 & 66.84 & 147.54 & 38.70 & 31.89 & 69.75 & 1.83 & 1.59 & 3.18\\
$t\bar t$  & 2109.00 & 754.80 & 2331.00 & 710.4 & 222.00 & 466.2 & 0.00 & 0.00 & 0.00\\
$Zb\bar b$ & 0.00 & 6.99 & 19.38 & 0.00 & 0.00 & 1.62 & 0.000 & 0.000 & 0.000\\
 \hline 
\end{tabular}
\end{center}
\caption{The number of events for $30~{\rm fb}^{-1}$ luminosity at $\sqrt s=14$ TeV LHC for the SSD+$nj+\mET$ signal (with $n\geq 2$) and the dominant SM backgrounds. We have shown the numbers after the basic selection criteria (but before applying any additional cuts) as well as after applying the following additional cuts: (i) $p_T^{\rm all~jets}>50$ GeV with $p_T^{\rm leading~jet}>100$ GeV, and (ii) $\mET>300$ GeV.}
\label{tab:signal}
\end{table} 

\begin{figure}[h!]
\centering
\includegraphics[width=6cm]{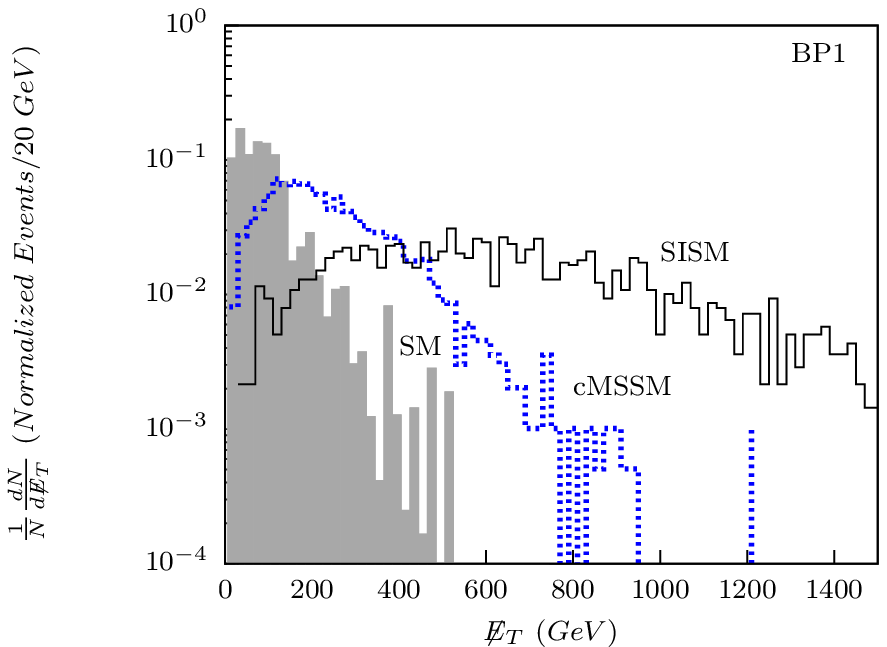}
\hspace{1cm}
\includegraphics[width=6cm]{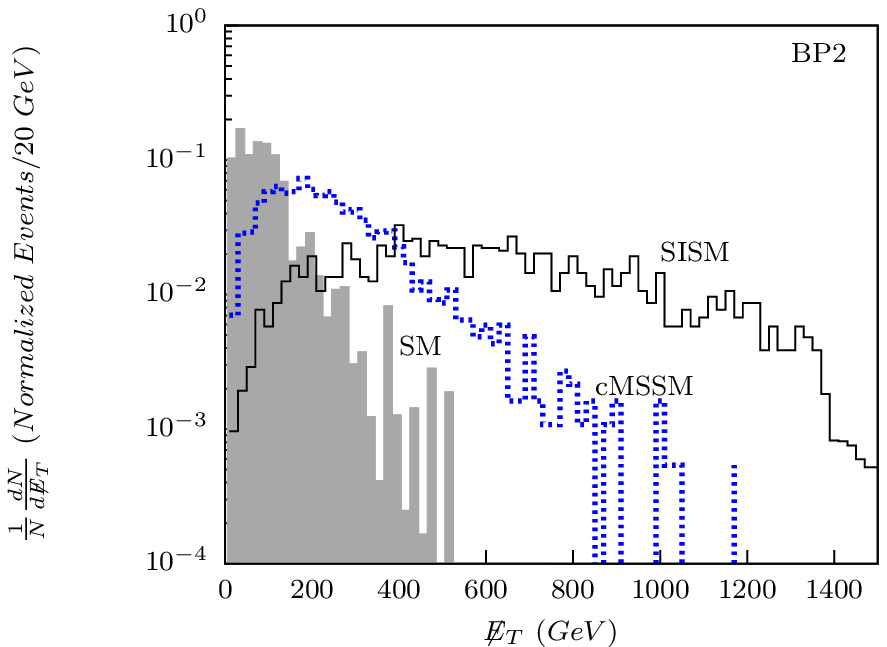}
\hspace{1cm}
\includegraphics[width=6cm]{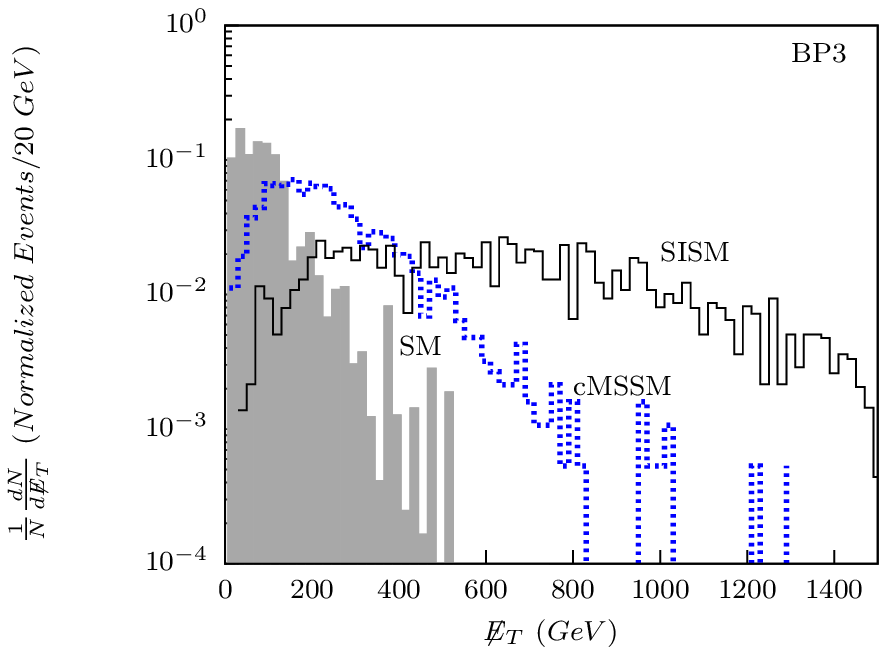}
\caption{The $\mET$~distribution for the SISM case with sneutrino LSP (solid lines) and the cMSSM case with neutralino LSP (dashed lines) with a similar squark-gluino spectrum. The SM background is also shown (shaded region).}
\label{fig:met}
\end{figure}

The $\mET$ distributions for both SISM and cMSSM cases are shown in Figure~\ref{fig:met} for all the benchmark points. It is clear that the SISM case has a much harder $\mET$ ~tail compared to the cMSSM case which can be used as a distinguishing feature. The combined SM background is also shown (in shades) which falls rapidly for $\mET>300$ GeV. This justifies our $\mET$ ~cut selection in Table~\ref{tab:signal}.   
\section{Summary and Conclusions}
We have considered a SUSY scenario that accommodates the inverse seesaw mechanism of neutrino mass generation via a small $\Delta L=2$ mass term (of the order $\sim$ keV). Two sets of $SU(2)_L$-singlet neutrino superfields have been introduced for this purpose. We show that this model can not only account for the neutrino masses and mixing, but also leads to an LSP dominated by right chiral sneutrino states. For phenomenologically consistent input parameters, taken as a hybrid of the top-down and bottom-up choices, the sneutrino LSP can act as a light DM candidate of mass around 50 GeV while satisfying all the existing collider, cosmological as well as low-energy constraints. We also suggest that such a scenario can be distinguished from one based on the usual mSUGRA scenario 
with a neutralino LSP, through a study of the same-sign dilepton signals at the LHC, and also from the $\mET$~ spectra in the two cases. We might also be 
able to put useful bounds on the Dirac Yukawa coupling in such scenarios from the invisible decay width of the lightest neutral Higgs boson if this gets 
confirmed with more data at the LHC in near future. 
\section*{Acknowledgments} 
We thank Florian Staub for useful correspondence. BD would like to thank Haipeng An and Rabindra Mohapatra for helpful discussions and collaboration at an earlier stage. BD and BM acknowledge the hospitality of the Indian Association for the Cultivation of Science (IACS), Kolkata, 
where a major part of this work was carried out. BD also acknowledges the hospitality of the Harish-Chandra Research Institute (HRI), Allahabad, during the 
final stages of this work. 
This work of BD was supported in part by the Lancaster-Manchester-Sheffield 
Consortium for Fundamental Physics under STFC grant ST/J000418/1. 
BM's work was partially supported by funding available from the Department of Atomic Energy, Government of India, for the Regional Centre for Accelerator-based Particle Physics, HRI, Allahabad. SM wishes to thank the Department of Science and Technology, Government of India, for a Senior Research Fellowship. SR acknowledges the hospitality of the Physics Department and the Cluster
of Excellence for Fundamental Physics `Origin and Structure of the Universe' at the Technical University in Munich (TUM) where a part of this work was done.   

\end{document}